
\def\Version{ 2 } 




  %





\message{ Assuming 8.5" x 11" paper }    

\magnification=\magstep1	          

\raggedbottom

\overfullrule=0pt 

\parskip=9pt

%

\def\singlespace{\baselineskip=12pt}      
\def\sesquispace{\baselineskip=16pt}      



\font\openface=msbm10 at10pt
 %

\def\Reals         {{\hbox{\openface R}}}
\def\Complexes     {{\hbox{\openface C}}}
\def\Minkowski     {{\hbox{\openface M}}}




\def\Re  {\mathop{\rm Re}  \nolimits}    

\def\im  {\mathop{\rm im}  \nolimits}    
\def\image  {\mathop{\rm image} \nolimits}    

\def\kernel {\mathop{\rm kernel} \nolimits}






\def\implies{\Rightarrow}



\def\sqr#1#2{\vcenter{
  \hrule height.#2pt 
  \hbox{\vrule width.#2pt height#1pt 
        \kern#1pt 
        \vrule width.#2pt}
  \hrule height.#2pt}}


\def\dal{\mathop{\,\sqr{7}{5}\,}}
\def\block{\dal} 
 %


\def\lto{\mathop
        {\hbox{${\lower3.8pt\hbox{$<$}}\atop{\raise0.2pt\hbox{$\sim$}}$}}}
\def\gto{\mathop
        {\hbox{${\lower3.8pt\hbox{$>$}}\atop{\raise0.2pt\hbox{$\sim$}}$}}}


\def\half{{1 \over 2}}


\def\part{\subseteq}		




\def\to{\mathop\rightarrow}	

\def\orthogonal{\mathop\bot}

\def\ideq{\equiv}		

\def\less{\backslash}		

\def\tensor{\otimes}		

\def\interior #1 {{ \buildrel \circ \over #1 }} 
 %




\def\grad{\nabla}


\def\basisvector#1#2#3{
 \lower6pt\hbox{
  ${\buildrel{\displaystyle #1}\over{\scriptscriptstyle(#2)}}$}^#3}

\def\alfa{\alpha}


\def\hat{\widehat}		
\def\tilde{\widetilde}		

\def\bar{\overline}		


\newcount\tmpnum \newdimen\tmpdim
{\lccode`\?=`\p \lccode`\!=`\t  \lowercase{\gdef\ignorept#1?!{#1}}}

\edef\widecharS{\expandafter\ignorept\the\fontdimen1\textfont1}

\def\widebar#1{\futurelet\next\widebarA#1\widebarA}
\def\widebarA#1\widebarA{%
   \def\tmp{0}\ifcat\noexpand\next A\def\tmp{1}\fi
   \widebarE
   \ifdim\tmp pt=0pt \overline{#1}%
   \else {\mathpalette\widebarB{#1}}\fi
}
\def\widebarB#1#2{%
   \setbox0=\hbox{$#1\overline{#2}$}%
   \tmpdim=\tmp\ht0 \advance\tmpdim by-.4pt
   \tmpdim=\widecharS\tmpdim
   \kern\tmpdim\overline{\kern-\tmpdim#2}%
}

\def\widebarC#1#2 {\ifx#1\end \else 
   \ifx#1\next\def\tmp{#2}\widebarD 
   \else\expandafter\expandafter\expandafter\widebarC
   \fi\fi
}
\def\widebarD#1\end. {\fi\fi}

\def\widebarE{\widebarC A1.4 J1.2 L.6 O.8 T.5 U.7 V.3 W.1 Y.2 
   a.5 b.2 d1.1 h.5 i.5 k.5 l.3 m.4 n.4 o.6 p.4 r.5 t.4 v.7 w.7 x.8 y.8
   \alpha1 \beta1 \gamma.6 \delta.8 \epsilon.8 \varepsilon.8 \zeta.6 \eta.4
   \theta.8 \vartheta.8 \iota.5 \kappa.8 \lambda.5 \mu1 \nu.5 \xi.7 \pi.6
   \varpi.9 \rho1 \varrho1 \sigma.7 \varsigma.7 \tau.6 \upsilon.7 \phi1
   \varphi.6 \chi.7 \psi1 \omega.5 \cal1 \end. }

\def\bar{\widebar}




\fontdimen16\textfont2=2.5pt
\fontdimen17\textfont2=2.5pt
\fontdimen14\textfont2=4.5pt
\fontdimen13\textfont2=4.5pt


\let\miguu=\footnote
\def\footnote#1#2{{$\,$\parindent=9pt\baselineskip=13pt%
\miguu{#1}{#2\vskip -7truept}}}
 %

\def\linebreak{\hfil\break}
\def\lbr{\linebreak}
\def\pagebreak{\vfil\break}


\def\BulletItem #1 {\item{$\bullet$}{#1 }}
\def\bulletitem #1 {\BulletItem{#1}}




\def\REMARK{\noindent {\csmc Remark \ }}

\def\PROOF{\noindent {\csmc Proof \ }}
\def\PROBLEM{\noindent {\csmc Problem \ }}

\def\QUESTION{\noindent {\csmc Question \ }}

\def\LEMMA{\bigskip\noindent {\csmc Lemma }}

\def\PrintVersionNumber{
 \vskip -1 true in \medskip 
 \rightline{version \Version} 
 \vskip 0.3 true in \bigskip \bigskip}

\def\author#1 {\medskip\centerline{\it #1}\bigskip}

\def\address#1{\centerline{\it #1}\smallskip}

\def\furtheraddress#1{\centerline{\it and}\smallskip\centerline{\it #1}\smallskip}

\def\email#1{\smallskip\centerline{\it address for email: #1}} 

\def\AbstractBegins
{
 \singlespace                                        
 \bigskip\leftskip=1.5truecm\rightskip=1.5truecm     
 \centerline{\bf Abstract}
 \smallskip
 \noindent	
 } 
\def\AbstractEnds
{
 \bigskip\leftskip=0truecm\rightskip=0truecm       
 }

\def\section #1 {\bigskip\noindent{\headingfont #1 }\par\nobreak\smallskip\noindent}

\def\subsection #1 {\medskip\noindent{\subheadfont #1 }\par\nobreak\smallskip\noindent}
 %

\def\ReferencesBegin
{
 \singlespace					   
 \vskip 0.5truein
 \centerline           {\bf References}
 \par\nobreak
 \medskip
 \noindent
 \parindent=2pt
 \parskip=6pt			
 }
 %

\def\reference{\hangindent=1pc\hangafter=1} 

\def\ref{\reference}

\def\sepref{\parskip=4pt \par \hangindent=1pc\hangafter=0}
 %

\def\journaldata#1#2#3#4{{\it #1\/}\phantom{--}{\bf #2$\,$:} $\!$#3 (#4)}
 %

\def\eprint#1{{\tt #1}}

\def\arxiv#1{\hbox{\tt http://arXiv.org/abs/#1}}
 %


\def\webhome{{\tt http://www.pitp.ca/personal/rsorkin/}}
 %

 %



\def\webtilde{\lower2pt\hbox{${\widetilde{\phantom{m}}}$}}

 %

\def\hpf#1{\webhome{\tt{some.papers/}}}
 %

\def\hpfll#1{\webhome{\tt{lisp.library/}}}
 %


\font\titlefont=cmb10 scaled\magstep2 

\font\headingfont=cmb10 at 12pt
 %

\font\subheadfont=cmssi10 scaled\magstep1 
 %

\font\csmc=cmcsc10  


\input miniltx
\input graphicx  



\def\Caption#1{
\vskip 1cm
\vbox{
 \leftskip=1.5truecm\rightskip=1.5truecm     
 \singlespace                                
 \noindent #1
 \vskip .25in\leftskip=0truecm\rightskip=0truecm 
} 
 \vskip 0.01cm
 \sesquispace}


\def\fhat{\hat{\phi}}

\def\KG{\bullet}

\def\bra{\langle}
\def\ket{\rangle}

\def\fihat{\hat{\phi}}
\def\ahat{\hat{a}}

\def\wbar{\bar{w}}
\def\Wbar{\bar{W}}

\def\phat{\hat{p}}
\def\qhat{\hat{q}}

\def\nablaa{{\buildrel\longleftrightarrow\over\nabla}}             
\def\blockk{{\buildrel\longleftrightarrow\over\block}}             
\def\overdoublearrow#1 { {\buildrel\longleftrightarrow\over{#1}} } 

\def\PROOF{\medskip\noindent {\csmc Proof \ }}



\phantom{}


\PrintVersionNumber   



\sesquispace
\centerline{{\titlefont From Green Function to Quantum Field}\footnote{$^{^{\displaystyle\star}}$}%
{Published in \journaldata{International J. of Geometric Methods in Modern Physics}{14}{1740007}{2017}
}}

\bigskip


\singlespace			        

\author{Rafael D. Sorkin}
\address
 {Perimeter Institute, 31 Caroline Street North, Waterloo ON, N2L 2Y5 Canada}
\furtheraddress
 {Department of Physics, Syracuse University, Syracuse, NY 13244-1130, U.S.A.}
\email{rsorkin@perimeterinstitute.ca}

\AbstractBegins                              
A pedagogical introduction to the theory of a gaussian scalar field
which shows firstly, how the whole theory is encapsulated in the
Wightman function $W(x,y)=\bra\phi(x)\phi(y)\ket$ regarded abstractly as
a two-index tensor on the vector space of (spacetime) field
configurations, and secondly how one can arrive at $W(x,y)$ starting
from nothing but the retarded Green function $G(x,y)$.  Conceiving the
theory in this manner seems well suited to curved spacetimes and to
causal sets.  It makes it possible to provide a general spacetime region
with a distinguished ``vacuum'' or ``ground state'', and to recognize
some interesting formal relationships, including a general condition on
$W(x,y)$ expressing zero-entropy or ``purity''.

\bigskip
\noindent {\it\/Keywords and phrases\/}:  
  gaussian field, S-J vacuum, Peierls bracket, \linebreak 
  ground-state condition, purity criterion, entropy of Wightman function
\AbstractEnds                                

\bigskip


\sesquispace
\vskip -10pt


\section{1.~Introduction and setup}                                 
Since these lectures are intended as a pedagogical introduction to a
certain type of quantum field theory, I should begin by explaining in
what sense we will be understanding that concept.  The specific theory
in question will be that of a gaussian (real) scalar field, by which I
mean what is often referred to as a free scalar field in a gaussian (or
``quasi-free'') state.  Thus I am including a ``choice of vacuum'' in
the definition of the theory.
Although doing this is not unheard of (consider the Wightman axioms, for
example), it could seem unfamiliar from the ``algebraic'' perspective
according to which the essence of a quantum theory is simply a $\star$-algebra
of ``observables'', with no distinguished ``state'' being specified.

I have nevertheless chosen to proceed this way for two reasons. First of
all, if one takes a quantum dynamics to be given by a quantum measure
[1] or
decoherence functional  [2]
(as opposed to a propagator, for example), then
one erases any principled distinction between ``initial state'' and
``equation of motion''.

Secondly, and more importantly, we will be proceeding in a somewhat
unorthodox fashion.  Rather than start with the field equations, we will
base our construction solely on the retarded Green function and the
volume-element $\sqrt{-g}\,d^{\,4}x$ of the spacetime in which the field will
live, following an approach which arose in the course of defining a
quantum field theory on a causal set.  [3] [4]
The field-operators will then be
derived from the two-point correlation function, which in turn will be
derived from the Green function.  Such an approach builds in
``state-information'' from the very beginning, in the form of the
two-point correlation function or ``Wightman function'', $W(x,x')$.

In this connection it seems fitting to call to mind some other well
known situations in which knowledge of a ``vacuum'', or of a set of
favoured ``vacua'', plays an important role.  Such are Hawking radiation,
the Unruh effect, and inflationary scenarios (cf.  their attendant
``trans-Planckian'' worries).  In the same vein, one can try to identify
a distinguished state or vacuum which is ``natural to'' the early
universe, supposing the underlying metric to be known.  If this were
done, and if the resulting vacuum turned out to be, for example, thermal
with approximately scale-free fluctuations, one would have taken a step
toward explaining (with no appeal to inflation) the initial conditions
that gave birth to our part of spacetime.

\subsection {The kinematic framework for these lectures}            
We begin with a spacetime manifold $M$, either compact with boundary
$\partial{M}$, or non-compact without boundary, and 
we write $\interior M  = M \less \partial M\,$. 
To signify that $x\in M$ lies in $J^-(y)$, the causal past of $y\in M$,
we will write ``$x\prec y$'', and we will define for $a,b\in M$,
$J(a,b)=\{x\in M | a \prec x \prec b\}$.
We will assume that $M$ is globally
hyperbolic in a sense generalizing that employed in [5]. 
Namely, we assume 
that $M$ is causally locally convex,
that $J(a,b)$ is compact $\forall a,b \in M$, 
and in the case with boundary that 
$J(a,b)$ is disjoint from $\partial{M}$ when $a,b\in\interior M \,$.   
In addition, one might
want to impose some form of completeness on $M$, for example that every
geodesic that doesn't end on $\partial{M}$ extends to infinite affine
parameter.

Our dynamical variable will be a real scalar field $\phi=\phi(x)$ of
mass $m\ge0$, massive or massless but in any case free.  (For brevity, I
will sometimes write intermediate formulas only for the massless case.)

Global hyperbolicity implies that the retarded Green function
$$
      G(x,y)\ideq G^{ret}(x,y)
$$
exists and is uniquely defined.  For definiteness, we will take it to be
a biscalar (i.e. a density of weight 0 in both its arguments).  We have
then
$$
     (\block - m^2) G(x,y) = \  \hat{\delta}(x,y) \ 
     \ideq \ 
     {\delta^{(4)}(x-y) \over \sqrt{-g(y)}}
     \eqno(1)
$$

We will often employ matrix or tensor-index notation for $G$ and other
two-point functions, writing for example ``$G^{xy}$'' instead of
``$G(x,y)$''.  Then $\tilde{G}$  will denote the transpose of $G$ and
$\bar{G}$ its  (elementwise) complex conjugate:
$$
           \tilde{G}^{xy} = G^{yx} \ , \quad \bar{G}^{xy} = (G^{xy})^*
$$
We will denote hermitian conjugate (adjoint) with a star, and write
$A\,\natural\,B$ to signify that $AB=BA$.
Our Lorentzian-metric signature will be that of $(-+++)$.

\bigskip\bigskip
\let\parskipsave=\parskip
\parskip=2pt
\singlespace
\centerline {\bf Table of Contents}
\item{1.} {Introduction and setup}
\item{2.} {Green functions and the commutator function}
\item{3.} {Gaussian fields (from Wightman function to field operators and Fok vacuum)}
\item{4.} {The ``S-J vacuum'' (from Green function to Wightman function)}
\item{5.} {Stationary spacetimes and the harmonic oscillator}
\smallskip
\vbox{
\item{6.} {For which $W$ does the entropy vanish?}
\item{} \hskip -12pt Acknowledgements 
}
\parskip=\parskipsave
\sesquispace
\bigskip

\section{2.~Green functions and the commutator function}            
Recall that $G$, written without any qualifying superscript, represents the
{\it\/retarded\/} Green function $G^{ret}$.  Thus $G(x,y)=0$ unless $x\succ y$.
In what follows it will important to know that the
transpose function $\tilde{G}$ is {\it\/also\/} a Green function, which by
definition is equivalent to it being the advanced Green function:
$$
    \tilde{G^{ret}} = G^{adv}           \eqno(2)
$$
In flat spacetime this equality holds trivially since 
$G(x,y)=G(x-y)$ by translation-symmetry, 
but we do not want to limit ourselves to that case.  
In the general case,
(2) can be proven as follows (see figure~1).

\PROOF 
First notice that, just as global hyperbolicity implies the existence and
uniqueness of the retarded Green function, it does the same for the advanced Green function.
Let $H=G^{adv}$ be this function and consider the integral
$$\eqalign{
     &\int G(x,a) \  \overdoublearrow{{(\block(x) - m^2)}} \ H(x,b) \; dV(x) \cr  
     =\  &\int G(x,a)\  \blockk(x)  \ H(x,b) \; dV(x)    
}
$$
where the double arrow indicates antisymmetrization as in the equality, 
 $A\nablaa B = A(\nabla B) - (\nabla A)B$.
On one hand (cf. (1)) our integral equals
$$\eqalign{
     &\int G(x,a) \  \hat\delta(x,b) \; dV(x)  - \int \hat\delta(x,a) H(x,b) dV \cr 
     &= \ G(b,a) - H(a,b)
}
$$
On the other hand it equals, by Stokes theorem, 
$$
     \oint\limits_{\partial{M}}  G(x,a) \, \nablaa{}\!^\mu(x) \; H(x,b) \; dS_\mu(x) 
$$
Now for
$a,b \, \in \, \interior M $,
the last integral necessarily vanishes
because its integrand is zero unless $a\prec x\prec b$,
whereas the global hyperbolicity of $M$ implies that 
$J(a,b)$
cannot touch $\partial{M}$ (or reach infinity).
%
\vskip 1.1cm       
 \vbox{\bigskip
   \centerline {\includegraphics[scale=0.5]{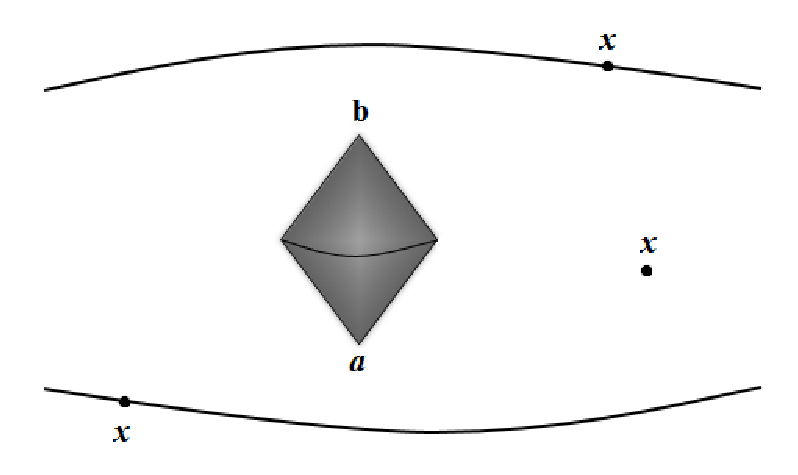}}
   \Caption{{\it Figure 1.}  Setup for the proof of (2).  The shaded region is $J(a,b)$.}}
\noindent
Then by continuity,
$$
     H(a,b) = G(b,a) = \tilde{G}(a,b)
$$
holds everywhere, as was to be proven.
Now let us define
$$
      \Delta = G - \tilde{G} \ , \eqno(3)
$$
noting then that
$(\block - m^2)\Delta=0$, and that
$\Delta$ is real and skew: $\tilde\Delta=-\Delta$.

The following lemmas are generally useful, and they will help to explain why our
construction of the field theory ``works'', however we won't need them as such
below.  For this reason 
I will state them only in a formal sense and will provide only a
rough proof.\footnote{$^\star$}
{The lemmas are only formal because we ignore niceties like taking the closure of
 $\im\Delta$ or characterizing the domain of $\Delta$ when $M$ is non-compact.}

\LEMMA 1.\ \   $ \image \Delta = \kernel (\block - m^2) $

\noindent
Before trying to prove this, let's state another lemma whose content is
essentially the same, but whose demonstration is slightly more direct
(see figures 2 and 3).

\LEMMA 2.\ \  $  \ker\Delta \  \orthogonal\  \ker(\block - m^2) $

In verifying this lemma, 
let us take $m=0$ for convenience, 
and further let $f\in\ker\Delta$ and
$\varphi\in\ker\block$.  We are then required to show
that $\int f \varphi dV=0$.
To that end, notice first that
$f\in \ker\Delta \iff \Delta f = 0 \iff Gf=\tilde{G}f \ideq h$.
I claim that  it follows from this that both $h$ and its gradient vanish on $\partial{M}$.
Accepting this for now, notice also that $\block h = \block G f = f$, 
since $\block{}G=1$ by definition of a Green function. 
Therefore,
$$
\eqalign{
   \int \varphi f dV &= \int \varphi \block h dV                    \cr
          &= \int \varphi \blockk h dV + \int (\block \varphi) h dV \cr
          &= \oint\limits_{\partial M} \varphi \nablaa h  \cdot dS  \cr
          &=0
}
$$
where we used $\block\varphi=0$ in the second line, and then in the fourth line
the fact that both $h$ and $\grad h$ vanish on $\partial{M}$.

 \vbox{\bigskip
   \centerline {\includegraphics[scale=0.5]{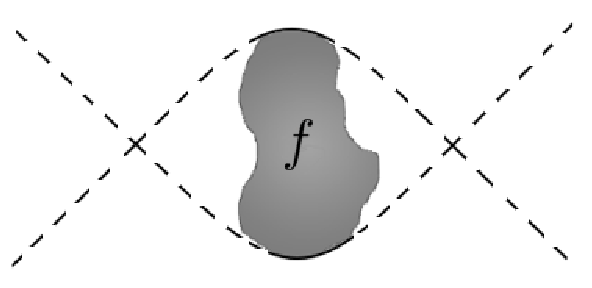}}
   \Caption{{\it Figure 2.}   Illustrating the support of $f$, its past and future}}

To complete the demonstration, we need to prove this last fact.  To that end,
observe first that our definition of global hyperbolicity precludes that any
finite portion of $\partial{M}$ is timelike.  For simplicity, let's also ignore any
null portions as well, and consider, say, $x\in\partial^-M$, the past boundary
of $M$.
It's then almost obvious (see figure 3) 
\vskip 1.1cm       
 \vbox{\bigskip
   \centerline {\includegraphics[scale=0.5]{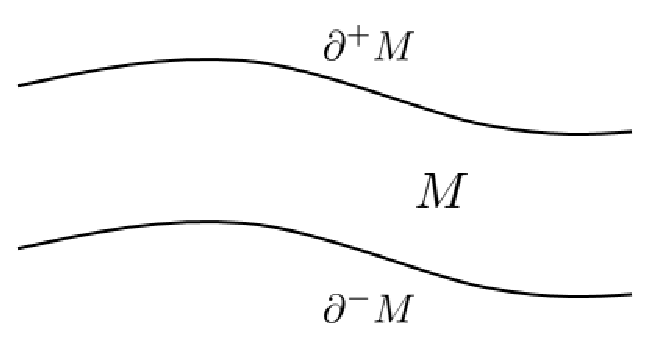}}
   \Caption{{\it Figure 3.}   The spacetime $M$ with its past and future boundaries}}

\noindent
that 
$h(x)=\int G(x,y)f(y)=0$ since only $y$ to the past of $x$ could have
contributed.  That $\grad h(x)$ also vanishes is less obvious, but it follows
from the observation that, on dimensional grounds the integral giving $h(x)$
must vanish like $\tau^2$ as $x \, \to\, \partial^{-}M$, provided that $f$ is bounded.


We can corroborate this dependence explicitly in both two and four
dimensions.  In the former case (and for $m^2=0$), $G=-1/2$ within the past
lightcone (and of course $G=0$ elsewhere), whence $\int G f$ will scale like
the area of $J^-(x)$, which in turn scales like $\tau^2$ as $\tau\to0$.
In the latter case the corresponding 4D integral looks like
$\int\limits_0^\tau dt \delta(r-t)/r \, r^2 dr \sim \int\limits_0^\tau t \,
dt \sim\tau^2$, once again.


\bigskip
\PROBLEM Clean up this proof.
\bigskip

With the aid of Lemma~2, we can easily complete the proof of Lemma~1.  
What we've shown so far is
$\ker\block \subseteq \orthogonal\ker\Delta = \im \tilde\Delta = \im\Delta$.
On the other hand, 
$\im\Delta \subseteq \ker\block$ trivially since $\block\Delta=0$.
Taken together these say that $\im\Delta=\ker\block$, as asserted.

\bigskip
\PROBLEM  Can we relate $\ker\Delta$ to $\im\block$?  
\smallskip\noindent
(Lemma~2 relates it to $\im\tilde{\block}$, but $\tilde{\block}$ is not
obviously the same as $\block$.)

\medskip
\REMARK  Lemma~1 implies that 
the image of $\Delta$ consists entirely of
solutions to the wave equation, $\block\phi=0$.
In this
sense the kernel (``nullspace'') of $\Delta$,
being orthogonal to its image,
has to be enormous, because the solutions $\phi$
are ``very few'' in relation to the set of all functions $\phi$ on $M$.  In
other words, $\im\Delta$ is ``very small'', while $\ker\Delta$ is ``very big''.
In the case of a causal set, 
this relationship would be meaningful without the quotation marks, 
because the dimensions of both $\im\Delta$ and $\ker\Delta$ would be finite,
but it
usually
turns out to hold only in an approximate manner:
$\Delta$ has very many tiny eigenvalues, but only a handful of
{\it\/exact\/}  $\,0$-modes!  One might say
in such cases
that the equation of motion for $\phi$ 
holds only approximately in the causal set.  
It turns out that this complicates the
definition of entanglement entropy in a way that is discussed further in 
[6].

\subsection {The commutation relations in covariant form (Peierls bracket)}  
With a free field, the commutator of any two field-operators $\fhat(x)$ is a $c$-number,
and we have the luxury of being able to express it in a manifestly covariant
form known as the Peierls bracket, namely as $[\fhat,\fhat]=i\Delta$, or more
explicitly:
$$
    [\fhat(x),\fhat(y)] = i\Delta(x,y)   \eqno(4)
$$

In order to relate this form to the more traditional equal-time commutation
relations, let us first examine a simple example in $\Minkowski^2$ with Cartesian
coordinates $t=x^0$ and $x=x^1$.  If, starting from the covariant form
$[\fhat(x),\fhat(y)] = i\Delta(x-y)$ with $y=0$, we take the derivative
$\partial/\partial{t}$ and then set $t=0$, we obtain
$$
      [\dot{\fhat}(t,x),\fhat(0,0)] = i \dot\Delta(t,x) |_{t=0} \ ,
$$ 
which we want to equal $-\delta(x)$.  Equivalently, we want
$$
         {\partial\Delta \over \partial t} |_{t=0} = - \delta(x) \ .
$$ 
But we know (defining $u=t-x$, $v=t+x$) that, as illustrated in figure~4,
$$
     G(u,v) = -\half \theta(u)\theta(v)
$$
 \vskip 1.1cm       
 \vbox{\bigskip
   \centerline {\includegraphics[scale=0.5]{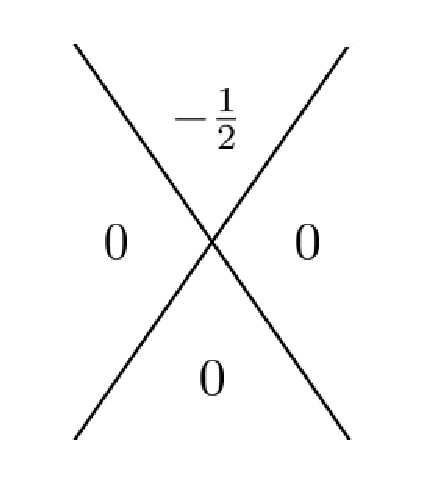}}
   \Caption{{\it Figure 4.} The massless Green-function $G(x)$ for $\Minkowski^2$}}
\noindent
whence 
$-2\Delta  = \theta(u)\theta(v) - \theta(-u)\theta(-v)$.
Noticing also that $\dot u = \dot v = 1$ and that at $t=0$,
$\delta(u)=\delta(v)=\delta(x)$, we easily compute
$$
\eqalign{
   -2 {\partial\Delta\over\partial t} &= \delta(u)\theta(v)+\theta(u)\delta(v)+\delta(u)\theta(-v)+\theta(-u)\delta(v)\cr
    &=\delta(u) [ \theta(v)+\theta(-v)] + (u\leftrightarrow v) \cr
    &=\delta(u)+\delta(v) \cr
    &= 2 \delta(x) \ \ [t=0] \cr
    {\partial\Delta\over\partial t} &= -\delta(x)
  }
$$
as required.

The general case is best handled in a coordinate-free fashion with the
aid of the Klein-Gordon inner product, defined for solutions $f$ and $g$
of the wave equation by\footnote{$^\dagger$}
{In the lectures, I denoted this inner product by ``KG'' with a circle
 around it, but I don't know how to make such symbol in plain Tex.}
$$ 
   f \KG g = \int\limits_\Sigma f  \nablaa{}\!^\mu g \; dS_\mu 
$$ 
where $\Sigma$, 
which can be any Cauchy surface, is externally oriented toward
the future (i.e. the sign of $dS_\mu$ is chosen to agree with that of
$\partial_\mu t$, where $t$ is chosen to vanish on $\Sigma$ and be
positive in the future of $\Sigma$, see figure 5).

 \vskip 1.1cm       
  \vbox{\bigskip
    \centerline {\includegraphics[scale=0.5]{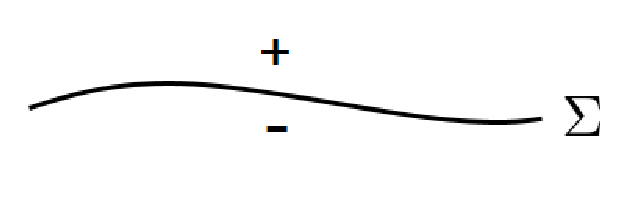}}
    \Caption{{\it Figure 5.}  Orientation of the surface-element $dS_\mu$ for the Klein-Gordon-inner product}}

\LEMMA 3.\ \ $(\block - m^2)f=0$  \ \ $\implies$\ \  $\Delta \KG f = f$

\PROOF (see figure 6)  Written out in full, the equation we are trying to prove says, at $x\in M$, 
$$ 
   \int\limits_{y\in\Sigma} \Delta(x,y) \nablaa{}\!^\mu(y) \, f(y) \, dS_\mu(y) = f(x) 
$$ 
Without loss of generality we can assume that $\Sigma\prec x \prec \Sigma'$,
where $\Sigma'$ is a second Cauchy surface introduced to the future of $x$, as shown.  
Then for $y\in\Sigma$,
$\Delta(x,y)=G(x,y)-G(y,x)=G(x,y)$, 
whence the integral we are evaluating becomes
$$
\eqalign{
   & \int\limits_{y\in\Sigma} G(x,y) \nablaa{}\!^\mu(y) \, f(y) \, dS_\mu(y) \cr
   &=
   - \big( \int\limits_{y\in\Sigma'} - \int\limits_{y\in\Sigma}  \big) \ 
    G(x,y) \nablaa{}\!^\mu(y) \, f(y) \, dS_\mu(y) \qquad ({\rm  since } \; \Sigma'\succ x) \cr
  &=
  - \int\limits_{\Sigma}^{\Sigma'} G(x,y) \blockk(y) f(y) \, dV(y) \qquad (\hbox{Stokes theorem})\cr
  &=
  \int\limits_{\Sigma}^{\Sigma'} \hat\delta(x,y) \,  f(y) \, dV(y) \cr
  &= f(x) 
} 
$$
(The minus sign in the third line arises because the Stokes theorem requires that $dS_\mu$ 
be oriented {\it\/outward\/}, 
which in the case of $\Sigma$ means pastward,
given that $\Sigma$ constitutes the past boundary of 
the region between $\Sigma$ and $\Sigma'$. [7] 
Notice in this connection that the ``boundary at spatial infinity'', if any,
does not contribute, because $G(x,y)$ vanishes outside some finite radius.)
This completes the proof of the lemma.

  \vbox{\bigskip
    \centerline {\includegraphics[scale=0.5]{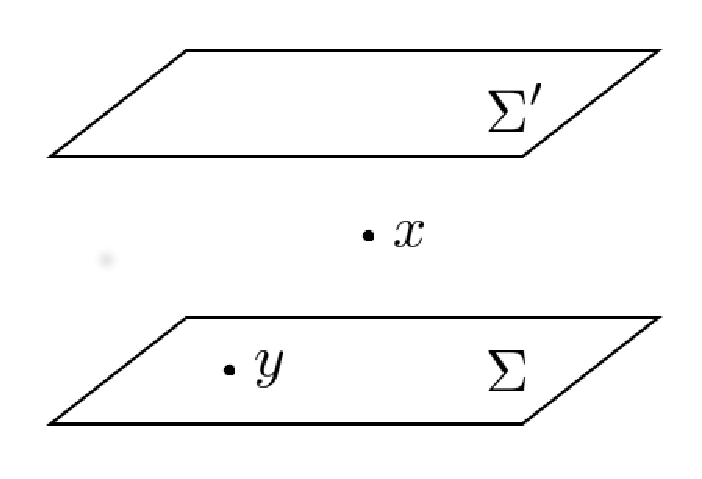}}
    \Caption{{\it Figure 6.}   Setup for proof of Lemma~3}}

Now let $f$ and $g$ again be solutions and consider the expression
$$ 
       f \KG \Delta \KG g  \eqno(5) 
$$ 
Evaluating it first directly and second with the aid of the lemma, we can
demonstrate that  (4) is equivalent to the equal-time commutation relations
that one commonly sees.  
In preparation, recall first that the surface-element
$dS_\mu(x)$ equals $n_\mu(x) d^3V(x)$, where $n_\mu$ is the unit
normal to $\Sigma$ at $x$ and $d^3V(x)$ is the volume-element of $\Sigma$.
And let a dot denote normal time-derivative, 
so that 
given our sign conventions,
$\dot{F}=-n^\mu\grad_\mu F$
because $n^\mu$ points pastward,
when $n_\mu$ has the same sign as $\partial_\mu{t}$.
Let us also make the ansatz that on  $\Sigma$, $\dot{f}=0$ and $g=0$.
(This ansatz is always possible, since any choice of $f$ and $\dot{f}$ yields
valid Cauchy data for the wave equation.)
For (5) we find then
$$
\eqalign{
   \int\int\limits_\Sigma &
     f(x) \; \nablaa{}\!^\mu(x) \; \Delta(x,y) \; \nablaa{}\!^\nu(y) \; g(y)  \ n_\mu(x) d^3V(x)\  n_\nu(y) d^3V(y) \cr
     &= \int\int f(x) \dot\Delta(x,y) \dot g(y) d^3V(x) d^3V(y)
}$$
On the other hand, the lemma teaches us that (5) is equal to $f \KG g$
$$\eqalign{
  &= \int f \; \nablaa{}\!_\mu \; g \   n^\mu  d^3V \cr
  &= \int (- f \dot g + \dot f g) d^3V \cr
  &= \int - f \dot g \, d^3V \cr
  &= - \int f(x) \; \hat\delta^{(3)}(x,y) \; \dot{g}(y) \  d^3V(x) d^3V(y)
}$$ 
Comparing the two expressions then reveals that 
on $\Sigma$, $\dot\Delta(x,y)=-\hat\delta^{(3)}(x,y)$.
(Recall that we defined $\hat\delta(x,y)$ to be a bi-scalar.)

We have thus established 
the nontrivial one of the three equal time commutation relations, the
counterpart of $[q,p]=i$.  The other two may be left as an exercise:

\PROBLEM  Show that the equal-time commutation relations $[\phi,\phi]=0$ and
$[\dot\phi,\dot\phi]=0$ follow similarly if one chooses $f=g=0$, respectively 
$\dot f=\dot g=0$.
\bigskip


\REMARK
Although the Klein-Gordon inner product will not be needed again (having
been introduced only to help us relate the Peierls bracket (4) to the
more familiar equal-time commutation relations),
it features so prominently in other presentations of field theory that
it seems worth drawing attention here to its close relationship with the
operator $\Delta$.  To that end, recall first that the Klein-Gordon
product is defined only between solutions of the wave equation,
${(\block - m^2)} f = 0$.  But for such solutions, $f$ and $g$,
Lemma~1 implies that $\Delta^{-1} g$ exists and is unique up to an
element $k\in\ker\Delta$, while Lemma~2 implies that any such $k$ is
orthogonal to $f$.  Thus, the two lemmas imply that the formula,
$f\Delta^{-1}g = f^x \Delta^{-1}_{xy}\, g^y$, evaluates to a well-defined
numerical value, and this value turns out to be none other than $f\KG
g$, as a derivation very similar to the proof of Lemma~3 will reveal.
In this way the Klein-Gordon inner product acquires a global or
``spacetime'' expression that is no longer tied so closely to the continuum.
(Being global, the same expression ought to carry over to the causet
case, provided that a serviceable form of the Klein-Gordon equation is
available there.)

\PROBLEM  Prove that \ $f \, \Delta^{-1}g = f \KG g$ \ for any two solutions $f$ and $g$
 
\bigskip

As stated earlier, we will build up our scalar field $\hat\phi$ in a manner
suggested originally by the needs of causal set theory, where certain features
of the (local) continuum theory seem to be available only in approximate form.
Essential to our work will be the recognition that one can characterize a
gaussian field with no other input than a two-point function $W(x,x')$, and it
is to this remarkable characteristic of gaussian theories that we now turn.


\section{3.~Gaussian fields (from $W(x,y)$ to a full field theory)} 
What, at a minimum, does it take to define a field theory?  If the field in
question is {\it\/gaussian\/}, then it suffices to have the one- and two-point
correlation functions, 
$$
    \bra \phi(x) \ket \qquad \hbox{and} \qquad W(x,y)=\bra \phi(x) \phi(y) \ket
$$ 
Or if, as is most often the case, 
and as we will assume herein,
$$
  \bra\phi(x)\ket=0  \ ,
$$ 
one can make do with
$W(x,y)$ alone.   Once this ``Wightman function''  is given, the remaining
$n$-point functions follow via Wick's rule:
$$
   \bra \phi(x_1) \phi(x_2) \cdots \phi(x_{n}) \ket 
   = \sum \bra \phi \phi \ket
   \bra \phi \phi\ket \cdots \bra \phi \phi \ket
   \eqno(6)
$$
where the sum is taken over all pairings of the arguments,
$\phi(x_1) \cdots \phi(x_n)$.  (In accord with the rule, the $n$-point
functions with $n$ odd all vanish.)

For present purposes we may take (6) as the meaning of gaussianity, inasmuch as
in the analogous situation with probability distributions, this ``Wickian'' pattern of
moments is known to characterize the gaussian distributions as a family, and individually
to determine any such distribution fully (compare also the Wightman axioms
for flat-space quantum field theory).
An analogous relationship for quantum field theory may be found in [8],
where it is shown how Wick's rule leads to a path-integral (more precisely a
{\it\/quantum-measure\/}) with a manifestly gaussian form.
Notice here that, as advertised in the introduction, we are regarding the
expectation-value functional, $\bra\cdot\ket$, as part of the definition of our
theory, which is why I have been speaking of a ``gaussian theory'' or ``gaussian
field'', and not for example, a ``gaussian state''.

Although it is true that $W$ determines the theory, it far from true that it can
be specified freely.  On the contrary $W(x,y)$ needs to satisfy the condition that,
regarded as a matrix --- or better as a quadratic form --- it is positive-semidefinite:\footnote{$^\flat$}
{If $f$ and $g$ are taken to be scalar densities rather than scalars then the
volume-element $dV$ will not appear in this condition.}
$$
   (\forall f) \quad \int  dV(x) \, f^*(x) \, W(x,y) \, f(y) \, dV(y) \ \ge \ 0 \eqno(7)
$$
This condition (or rather the attendant condition 
of ``strong positivity of the quantum measure'') is perhaps best
regarded as an irreducible axiom of quantum theory.
However it can also be derived as a theorem, provided that one knows
that $\bra\cdot\ket$ can be expressed as expectation value with respect
to a state-vector $|0\ket\;$ (or a density-matrix) in some Hilbert space.
This follows immediately from the positivity of
$||\psi||^2=\bra\psi|\psi\ket$ with $\psi=\int dV(x) f(x) \fihat(x)|0\ket$.
Among its other consequences, (7) guarantees, via the
fluctuation-dissipation theorem, the ``passivity'' condition that one cannot
extract work from a system in thermal equilibrium by purely mechanical means
(cf. [9]).

Granted positivity, we can obtain a Fok space and field operators $\fihat(x)$
thereon simply by diagonalizing $W$ and introducing the corresponding
``annihilation operators'' $\ahat$, as we will see in a moment.  By doing so, we
can verify that nothing is missing from our formulation that one might want to
include under the rubric of ``field theory''.  

First however, let me introduce a notation and a viewpoint that will
allow us to conduct the passage from $W$ to $\fihat$ in a more abstract
and general setting.
For us so far, a field has been a function $f:M\to\Reals$, but if we
abstract from this particular context, we can view it simply as an
element of a real vector-space $V$, and accordingly view $W$ as a tensor
in $V\tensor V\tensor\Complexes$.

To diagonalize the tensor $W$ means to cast it into the form
$$
   W = \sum\limits_k  w_k\tensor \wbar_k    \eqno(8)
$$
where the $w$ are vectors in $V\tensor\Complexes$. 
Written in an indicial notation, (8) reads 
$W^{xy}=\sum\limits_k  w_k^x \, \wbar_k^y$, which of course would be
interpreted in our specific context as
$W(x,y)=\sum\limits_k  w_k(x) \, \wbar_k(y)$. \   
(Because $W$ is positive, all of the terms in (8) necessarily
 take the form $+w\tensor\wbar$ and none the form $-w\tensor\wbar$.)
On the basis of (8)
we can introduce in the familiar manner,
ladder-operators $\ahat_k$ 
such that 
$$ 
   [\, \ahat_k , \ahat_l^{\,*} \,] = \delta_{kl} \qquad \hbox{(say)}  \eqno(9)
$$ 
and a ``Fok vacuum'' $|0\ket$ 
such that $\ahat_k|0\ket=0$,
defining then 
$$ 
   \fihat = \sum\limits_k (w_k\,\ahat_k + \wbar_k\,\ahat_k^{\,*}) \eqno(10)
$$ 

That these definitions correctly reproduce $W$ is not hard to show.  Indeed, with
$\bra\cdot\ket$ taken to be $\bra0|\cdot|0\ket$, we have
$$\eqalign{ 
  \bra \fihat \tensor \fihat \ket
  &= \sum\limits_k  \bra (w_k \ahat_k+\wbar_k\ahat_k^*) \tensor (w_l \ahat_l+\wbar_l\ahat_l^*)\ket \cr
  &= \sum\limits_k   (w_k\tensor\wbar_l \bra \ahat_k \ahat_l^*\ket +0 + 0 + 0) \cr
  &= \sum\limits_k  w_k\tensor\wbar_k \cr
 &= W
}$$ 
It is equally straightforward to demonstrate 
the ``canonical commutation relations'' in their abstract form:
$$
    [\,\fihat^{\,x},\; \fihat^{\,y}\,] = W^{xy}-W^{yx}  \eqno(11)
$$
[The expectation value of (11) follows immediately from the
 calculation just above, but to derive it as relation among operators
 requires a separate proof.  Notice also that we have not written the
 right hand side of (11) as $i\Delta^{xy}$.  We could have done so,
 but in the present development, that would count as the definition of
 $\Delta$, not a relation between $W$ and some independently defined
 object.]

\bigskip
\PROBLEM\  Equation (11) can also be written as \  $\fihat\wedge\fihat=W-\bar{W}$, where 
 $a\wedge b \ideq a\tensor b - b\tensor a$.
 Prove it in this form, with the aid of (8) and (9).
\bigskip

In (10), we have written a real (selfadjoint) operator $\fihat$ in
terms of complex operators $\ahat$ and complex vectors $w$.  One could
also work exclusively with real operators $\qhat$ and $\phat$ for which
$[\qhat,\phat]=i$, 
and real vectors $u$ and $v$ for which
$$ 
   W = \sum \half (u\tensor u + v\tensor v + i u\wedge v) \eqno(12) 
$$ 
The ``purely real'' form of (10) is then 
$$
    \fihat = \sum \; (u \, \qhat + v\, \phat\,) \ , \eqno(13) 
$$  
the relation between the two representations being given by
 $w=(u-iv)/\sqrt{2}$ and $\ahat=(\qhat+i\phat\,)/\sqrt{2}$.
Note however that the set of $u$'s and $v$'s defined this way need not
be linearly independent, even though the set of $w$'s defined by (8) is.

The construction of vacuum-vector and field-operators from $W$ is now complete, but a
few further comments are in order, before returning to the more concrete context
where our field $\phi$ is a function on the manifold $M$.

A question that might arise in relation to the decomposition (8)
is why we didn't bother to normalize the vectors $w_k$.  The answer in this case
is that the question presupposes some reference metric with respect to which
such a normalization could take place.  In this section, we have been aiming at
generality, and therefore have taken care to avoid introducing any metric on
$V$.  When $V$ consists of functions on $M$, however, a metric is available
in the form of the $L^2$-inner product, 
and we will in fact use it in the next section 
to formulate a certain ``ground-state condition'' 
that can serve to determine $W$ uniquely in certain cases.
Here, though, we
have simply taken $W$ as given, without asking where it might have come from.

A second question concerns the uniqueness of the Fok representation we have just
set up.  The only input was supposed to be $W$, but what would have happened had
we used a different set of vectors $w_k$ to diagonalize $W$?  To this the answer
is that nothing would have happened.  Given that any second set of
vectors $w_k'$ will be related to the first by a unitary matrix 
(i.e.  $w_k'=\sum U_{kl}w_l$), 
it is not hard to see that one arrives thereby at a unitarily
equivalent Fok representation.  That is, our representation,
$\fihat$-cum-$|0\ket$, is unique up to unitary equivalence.

\bigskip
\PROBLEM  Show that if one defines the ``tensor operator'' $\ahat$ by
$$
       \ahat^{\,x} = \sum\limits_k w_k^x \  \ahat_k
$$ 
then our ansatzes for $\ahat_k$, $\fihat$, and $|0\ket$ 
result in equations from which any reference to the specific vectors
$w_k$ introduced to diagonalize $W$ has dropped out, to wit:
$$
\eqalign{
        & \fihat{\,^x} = \ahat{\,^x} + (\ahat{\,^x})^*  \cr
        & [\,\ahat{\,^x},\,(\ahat^{\,y})^*] = W^{xy}  \cr
        & \ahat{\,^x} |0\ket = 0
}
$$ 

A third question that one might ask is not disposed of so simply, because it
emphasizes something that we have {\it\/not\/} claimed about our construction.
What if the tensor $W$ from which we started were that of an ``impure'' gaussian
theory, such as that of a system in thermal equilibrium?  The entropy would then
have to be nonzero, whereas our Fok vacuum seems to be that of the
``pure state'' $|0\ket$.
The resolution to this paradox is that 
(as with the closely related case of GNS representations of *-algebras) 
our Fok vacuum can be impure
if the operators $\fihat$ defined by (10) 
fail to act irreducibly in the Fok space. 
Of course the operators $\,\ahat\,$ and $\,\ahat^{\,*}$ do act irreducibly, 
essentially by definition, 
but that does not necessarily imply that the $\fihat$ do so as well,
because one might not be able to recover the former from the latter.  This
{\it\/would\/} follow if, with respect to some hermitian reference metric, the
$w_k$ and the $\wbar_k$ were all orthogonal to each other (which actually will be
one of our ``ground-state conditions'' below), but it need not be true always.

That the $\ahat$ might not be recoverable from the $\fihat$ is perhaps easiest to
understand in finite dimensions, where the index on $\phi^x$ can be taken to run
from $1$ to $N$.  
The decisive question, then, is how many $w_k$ there are in
relation to the $\phi^x$.  Let $k$ run from $1$ to $M$, where $M$ is the
rank of $W$ as a matrix (the dimension of its image).  
From $N$ components $\fihat^x$, we need to recover $2M$
operators, $\ahat_k$ and $\ahat_k^*$, which is conceivable only when $2M\le N$.
Thus, purity (zero entropy) requires that rank($W$) be sufficiently small.
A  simple example is $N=2$, with $\phi^1=q$ and $\phi^2=p$, where purity
corresponds to the rank of $W$ being 1 rather than 2.


A final comment is that if $W$ happens to satisfy (exactly) some ``equation(s) of
motion'' (which means abstractly that $v_x W^{xy}=0$ for some covector or set of
covectors $v_k$), then plainly our field operator $\fihat$ will satisfy the same
equation(s).  But since we never needed any such equation of motion, our construction
will go through  equally well in the causal set case where such equations might hold
only approximately.  This of course was one motivation for proceeding in the
way we have.  

In the present section, we have raised our whole edifice on the foundation of a real
vector space $V$ and a  Wightman tensor $W\in V\tensor V\tensor\Complexes$.
But where does $W$ come from in practice?  One answer to this question is
provided by the so-called SJ-Ansatz, and it is to that which we turn next.

\section{4.~The ``S-J vacuum'' (from $G(x,y)$ to $W(x,y)$)}         
Returning to the setting of scalar functions on a globally hyperbolic spacetime, 
let us assume that we are given a function $G=G^{ret}$ to play the role
of the retarded Green function.  We will derive a distinguished Wightman function
$W$ from $G$ by way of $\Delta$.   In contrast, recall the usual story as
symbolized by the progression,
$$
   \block-m^2 \ \ \to\ \ G \ \ \to\ \  \Delta \ \ \to\ \  [\,\fihat\,,\,\fihat\,\,] 
   \quad {\buildrel \hbox{+ freq } \over {\,\longrightarrow\!\longrightarrow}}\quad  
   \ahat \ \ \to\ \ |0\ket \ \ \to\ \  W 
$$ 
Here we will tread the shorter path, 
$$  
         G \ \to\  \Delta \ \to\  W \ ,  
$$ 
in whose following out 
the Klein-Gordon field equation will play no role, 
except implicitly through its connection with the Green function $G$ from 
which $\Delta$ is formed.

To obtain $\Delta$ from $G$, we simply recall the Peierls prescription, 
$$ 
            \Delta = G - \tilde G  \ ,
$$  
which then furnishes information on $W$ itself via the relation,
$$ 
    W - \Wbar = i \Delta 
$$  
Note here that since $\tilde\Delta=-\Delta$ the tensor $i\Delta$ is antisymmetric
and pure imaginary, hence hermitian.
We also know that $W$ must be positive: $W\ge0$.  From this, $\Wbar\ge0$ 
follows as well, so that 
$W-\Wbar=i\Delta \implies \  W=i\Delta+\Wbar \ge i\Delta$.  
In a certain sense 
(relative to (14)),
our $W$ will be the smallest tensor which is $\ge i\Delta$

This is the basic idea, but the detailed prescription that implements it can be
given in three different, but equivalent, forms.\footnote{$^\star$}
{More background on the S-J vacuum can be found in
[10]
[4]
[11]
[12]
[13]
[14]
}
$L^2$-inner product on scalar fields (or more abstractly on the vector-space
$V$) will play an important part:
$$ 
        \bra f | g \ket = \int d^4V(x)\, f(x)^* \, g(x) \eqno(14)
$$ 
The $W$ that we will thereby build up has come to be known as that
of the ``S-J vacuum''.  As will become clearer in the following section, it
provides a kind of ``ground state'' that continues to be defined even in the
presence of curvature and the absence of any Killing vector.

\noindent{\bf First prescription.}  
By availing ourselves of the inner product that occurs in (14), we
can ``lower one of the indices'' on $\Delta$, and thereby construe it as
an operator from $V\tensor\Complexes$ to itself. This in turn makes it
possible to form powers and more general functions of $\Delta$, and we
use this ability in stating our first prescription for $W$:
$$
      W = \hbox{Pos}(i\Delta) = {i\Delta + \sqrt{-\Delta^2} \over 2} \eqno(15)
$$ 
Clearly $W\ge0$,  
and we have $W=R+i\Delta/2$ with $R=\half\sqrt{-\Delta^2}$.
Hence $W - \Wbar = i \Delta$ as required.

\REMARK  If we can compute $\sqrt{-\Delta^2}$ directly, then we never need to
diagonalize anything!  

\PROBLEM  Devise an algorithm to compute $\sqrt{-\Delta^2}$ without having to
diagonalize $i\Delta$.   Potential methods: resolvent? iterative? other?
\bigskip

\noindent{\bf Second prescription.}  
A more explicit expression for $W$ comes from diagonalizing $\Delta$.  
Partly for variety, but also because it is useful for computer arithmetic,
I will describe this in real form, 
where it amounts to doing a singular-value decomposition of $\Delta$, or
equivalently to block-diagonalizing the real skew matrix $\Delta$ into
$2\times2$-blocks of the form $\pmatrix{0 & \sigma \cr -\sigma & 0}$.  
The resulting form for $\Delta$ is 
$$
     \Delta = \sum\limits_j \sigma_j \ u_j \wedge v_j  \eqno(16)
$$ 
where the $\sigma_j>0$ are the so-called singular values, and $u_j$ and $v_j$ are real
vectors that together form an orthonormal family:
$$\eqalign{
           \int_M u_j u_k dV = \delta_{jk}    ,\quad
           \int_M u_j v_k dV = 0               ,\quad
           \int_M v_j v_k dV = \delta_{jk} 
}
\eqno(17)
$$
Written out fully, (16) reads as
$$ 
       \Delta(x,y) = \sum\limits_j \sigma_j \  (u_j(x) \, v_j(y) - v_j(x) \, u_j(y))
$$ 
Substituting (16) into (15) produces successively
$$
    - \Delta^2 = - \sum \sigma^2 \; (u\wedge v) \cdot (u\wedge v)
               = \sum \sigma^2 \; (u\tensor u + v\tensor v)
$$ 
$$ 
           \sqrt{- \Delta^2} = \sigma \; (u\tensor u + v\tensor v)
$$ 
$$ 
      W = \sum {\sigma\over2} (i u\wedge v + u\tensor u + v\tensor v)
        = \sum {\sigma\over2} (i u\tensor v - i v\tensor u + u\tensor u + v\tensor v)
   \eqno(18)
$$ 
Thus our second prescription furnishes $W$ in the form
$$ 
         W =   \sum \, \sigma \;  {u-iv \over \sqrt{2}}\tensor {u+iv \over \sqrt{2}}  
          \ , \eqno(19)
$$ 
which we can also write as $W=\sum\sigma \; w\tensor \wbar$ 
with $w_j=(u_j-iv_j)/\sqrt{2}$.

This last form is the special case of (8) where 
we choose the $w_k$ to diagonalize not only $W$ but also (14), 
and where we also normalize them with respect to (14).
If desired, one can of course go on to define operators $\fihat(x)$ from (19),
just as in (10) or (13).

Notice here that in  diagonalizing $W$, we have also diagonalized the hermitian
tensor $i\Delta$,
because in
$$
    i \Delta 
    = W - \Wbar 
    = \sum\limits_j \sigma_j \; w_j\tensor\wbar_j - \sigma_j \; \wbar_j\tensor w_j   \eqno(20) 
$$ 
the $w$ and $\wbar$ are orthogonal, thanks to (17).
The $w_j$ [resp. $\wbar_j$] 
are thus the 
eigenvectors
of $i\Delta$ with
positive [resp. negative] 
eigenvalue.
We could of course have started with this form, 
rather than with (16).

\medskip
\REMARK  In 4D with $M$ compact, one can show that $i\Delta$ is self-adjoint,
whence ``diagonalizable''. [13]

\noindent{\bf Third prescription.} From the first prescription, we have
$$
   4 \, W \, \Wbar 
             = (\sqrt{-\Delta^2}+i\Delta)(\sqrt{-\Delta^2}-i\Delta)
             = (\sqrt{-\Delta^2})^2 + \Delta^2 
             = - \Delta^2 + \Delta^2 = 0
$$
Since $W=W^*$, this is equivalent to $W\tilde{W}=0$, which in turn says that the
rows (and columns) of $W$ square to zero.  Still another way to say the same
thing is $W\orthogonal \Wbar$, meaning that $W$ and $\Wbar$ have disjoint supports.
(Notice that $W \Wbar=0 \implies \Wbar W=0 \implies W \; \natural\; \Wbar$, 
where `$\natural$' denotes ``commutes with''.)
These relationships make it possible to characterize $W$ in a third way, as
follows.  
$$ 
\eqalign{
  {(\alfa)} \ &{W - \Wbar = i\Delta} \cr
  {(\beta)} \ &{W \; \Wbar = 0}      \cr
  {(\gamma)}\ & {W\ge0}}
$$
Let us prove that $(\alfa)-(\gamma)$ determine $W$ uniquely, given $\Delta$.
\linebreak
$(\alfa) \implies W = R + i \Delta/2$ where $R=\bar R$
\linebreak
$(\beta) \implies (R + i \Delta/2)(R - i \Delta/2)=0$
\linebreak
$\phantom{(\beta)} \implies R^2 + \Delta^2/4 + i/2 [\Delta,R]=0$
\linebreak
$\phantom{(\beta)} \implies R^2 = - \Delta^2/4 \hbox{\qquad  \&  \qquad} \Delta \; \natural \; R$
\linebreak
$(\gamma) \implies R = \Re W \ge0 
          \implies R = \sqrt{-\Delta^2/4} \implies W = R+i\Delta/2 = (\sqrt{-\Delta^2}+i\Delta)/2$,
\linebreak
as required by (15).

In this third characterization of $W$, the crucial new condition is $(\beta)$.
We might therefore term it ``the ground state condition''.  In light of an
earlier remark, it would also make sense to regard it as a ``purity condition'',
but it seems best to reserve that exact phrase for a more general condition
we will encounter later, of which $(\beta)$ is a special case.

\bigskip
\REMARK  Our first two conditions, being equivalent to the quadratic
equations, $R^2=-\Delta^2/4$ and $R\Delta=\Delta R$, are relatively simple
algebraic requirements on the real matrices $R$ and $\Delta$.  
It seems probable that,
given $\Delta$, 
they by themselves determine $W$ up to some sign ambiguities that
$(\gamma)$ would then resolve as its only significant input to the prescription.
This suggests yet another possible way to avoid having to diagonalize any
matrices in building $W$.
\bigskip

In the previous section we learned that $W$ fully determines our gaussian theory,
and in particular that it furnishes an essentially unique field of operators
$\fihat(x)$.  To round out our story, let us prove that this field satisfies the
standard equation of motion, $(\block-m^2)\fihat=0$.

This follows first of all from $\ker(\block-m^2)=\im\Delta$,  
which is Lemma~1 proven in Section 2.
Combining it with (10), 
which shows that $\fihat(x)$ is assembled from vectors $w(x)$ in the image of $W$, 
and with (15), 
which shows that the image of $W$ is included in that of $\Delta$, 
we conclude, as desired, that $\fihat$ is assembled entirely
from solutions of the Klein-Gordon equation.

Alternatively, without invoking that lemma, we can simply note, as seen in 
(10) and (20), 
that $\fihat$ is built exclusively from eigenfunctions of $\Delta$ with strictly positive
eigenvalues.  Since such an eigenfunction $f$ is by definition a multiple of
$\Delta f$, it is annihilated by the wave operator, because $\Delta$ itself is.

\section{5.~Stationary spacetimes and the harmonic oscillator}      
Regarded as a $0+1$-dimensional scalar field, the simple harmonic oscillator is
the special case of our considerations given by $m^2=\omega^2>0$ and
$M=\Minkowski^1$, or $M=[0,T]\subseteq\Minkowski^1$ (where $\Minkowski^D$ is 
Minkowski spacetime of $D$ dimensions.)
Let us compute the S-J vacuum for this case.

In the present context the equation for the retarded Green function can be
written as
$$
  (\block-\omega^2)G = (-{\partial^2 \over \partial t^2} - \omega^2)G = \delta(t)
$$
Evidently $G(t)$ will be a linear combination 
of $\sin\omega t$ and $\cos\omega t$,
and it turns out that
$$
      G(t) = {-1\over\omega} \sin\omega t \; \theta(t) \ ,
$$ 
\PROOF
$\dot{G}=-\cos\omega t \theta(t) - (1/\omega) \sin\omega t \delta(t)$,
the second term of which vanishes inasmuch as $t\,\delta(t)=0\,\delta(t)=0$. 
Differentiating again yields
$\ddot{G} = \omega\sin(\omega t)\theta(t) - \cos(\omega t)\delta(t) = \omega\sin(\omega t)\theta(t) -\delta(t)$,
whereupon
$\ddot{G}+\omega^2 G = 0 - \delta(t)$
as required. \linebreak
From this follows immediately
$$
           \Delta(t) = G(t) - G(-t) = {-\sin(\omega t)\over \omega} \ ,
$$
the unique odd extension of $G(t)$.

We can now proceed in either ``real'' or ``complex'' mode.  Let's do the latter.\footnote{$^\dagger$}
{In real mode we'd deduce $u(t)$ and $v(t)$ directly from the decomposition
 $\sin\omega(t-t')=\sin\omega t\; \cos\omega t' - \cos\omega t\; \sin\omega t'$}
To that end, we need to cast $i\Delta(t,t')=i\Delta(t-t')$ into the mould of (20).
When $M=(-\infty,\infty)$  we do this almost unthinkingly, simply by writing
sine as a difference of exponentials,
producing thereby
$$
   i \Delta(t) = {1\over2\omega} (e^{-i\omega t} -  e^{i\omega t})
$$
$$
   i \Delta(t,t') = {1\over 2\omega} (e^{-i\omega t} e^{i\omega t'} - e^{i\omega t} e^{-i\omega t'}) \eqno(21)
$$
which has the form (20) with 
$$
    w(t) =  {1\over \sqrt{2\omega}} e^{-i\omega t}
$$
Moreover, on $M=(-\infty,\infty)$, 
this does in fact diagonalize $i\Delta$, 
because $\bra w|\wbar\ket=0$.  
Thus the positive part of $i\Delta$ 
demanded by (15)
is seen to be
$$
       W(t,t') = {1\over 2\omega}  e^{-i\omega t} e^{i\omega t'} = {e^{-i\omega (t-t')} \over 2\omega}
      \eqno(22)
$$ 
Notice that everything in this formula but the factor of $1/2$ follows from
the general properties of $W$.
(That $w$ cannot be normalized because it is not square integrable clearly does
not affect $W$ itself, which one obtains from (21) simply by crossing out
the negative term.)

Incidentally, does (22) satisfy our ground-state condition, which
required that the columns of $W$ all ``square to zero''?  Well, in this case, there
is one such column-vector for each $t'$, and all them are proportional to the
single function $f(t)=e^{-i\omega t}$.  But for this function,
$$
        \int\limits_\infty^\infty f(t)^2 dt 
       = \int\limits_\infty^\infty e^{-2i\omega t} dt = 2\pi\delta(2\omega) = 0
$$
since $\omega\not=0$.  The answer is therefore ``yes'' when we work on all of
$\Minkowski^1$, 
but notice that on $M=[0,T]$ we would have had instead,
$$
 \int\limits_0^T e^{-2i\omega t} dt \sim 1/\omega T     \eqno(23)
$$
which does not vanish unless $\omega T$ happens to be a multiple of $\pi$.
(Notice here, in passing, that even on all of $\Minkowski^1$, something like
$\sin\omega t$ or $\cos\omega t$ would not have worked as a ``row'' or ``column''
because its square would have integrated to infinity.  Notice also that
the ``squaring'' in question is purely algebraic, involving no complex conjugation:
the integrand was $f(t)f(t)$, not $f^*(t)f(t)$)

When $M=[0,t]$ is compact, it thus takes a bit more work to diagonalize $W$.
Instead of a single exponential, we will need 
$w(t)=\alfa e^{-i\omega t} + \beta e^{+i\omega t}$,
where $\beta/\alfa=O(1/\omega T)$ 
as follows from (23). 
For $\omega T\gg1$,
the resulting $W$ will therefore contain a small admixture
of negative frequencies, that is, the S-J vacuum in this case will
differ from the Minkowski vacuum by a small Bogoliubov transformation. 
The oscillator will be slightly excited.

The same thing will happen in higher dimensions when $M$ is compact, but
now it will happen ``mode by mode'', with the consequence that $W$ will
in general not have the so-called Hadamard form.\footnote{$^\flat$}
{See [13].  Although the ``occupation probability'' of any given
 ``mode'' will die out like $1/k^2$, as we just saw, the number of modes grows
 so rapidly in $D=4$ that the ``net occupation number'' will correspond to the
 divergent integral $\int k^2 dk/k^2$.  }
Since this form enjoys a special status in curved-space quantum field
theory, one might wonder whether the S-J prescription could be modified
to accommodate it.  

I believe that the answer is yes, and that one can
in fact do this rather easily, 
at least when $\partial{M}$ is everywhere spacelike,  
by modifying the spacetime volume-element, $dV=\sqrt{-g}\,d^4x$, 
in such a way as to ``soften the boundary of $M$'':
$$
    dV   \ \to\  \tilde{dV} = \rho(x) \;\sqrt{-g}\, d^4x  \ , \eqno(24) 
$$ 
where 
$\rho(x)\to0$ smoothly at $\partial{M}$ (see figure~7).
 \vskip 1.1cm       
 \vbox{\bigskip
   \centerline {\includegraphics[scale=0.5]{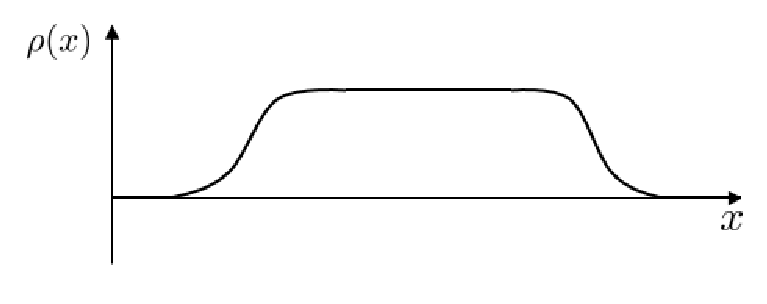}}
   \Caption{{\it Figure 7.}   A function $\rho$ that could be used to ``soften the boundary''}}
\noindent
For the harmonic oscillator, 
this will enhance the $O(1/\omega)$ falloff exhibited in (23)
to be faster than any power of the frequency (an essentially exponential falloff), 
and the same should happen mode by mode in 
more general spacetimes (not necessarily stationary). 
Indeed, a related technique 
(which however refers to an ambient spacetime in which $M$ is embedded) 
is developed in [15], 
and the resulting $W$ proven to be Hadamard in a range of cases. 

If the method based on (24) also works, 
as I believe it will, 
it will have the advantage of being ``self-contained''.  
Moreover, one can easily make a ``universal'' choice of $\rho$,  
by taking it to be a fixed function of 
some convenient ``distance to the boundary''. 
For example, 
letting $V^\pm$ denote the volume of $J^\pm(x)$,
one could take 
$\rho(x)=f(1/(1/V^+ + 1/V^-))$,
where $f(v)\ge0$ is a fixed function, 
chosen once and for all, 
that goes smoothly to 0 at $v=0$
and smoothly to 1 at some $v_0$ which sets the scale of the softening.
Once such an $f$ has been chosen,
(24) provides
for any compact $M$, 
a distinguished vacuum
whose UV behaviour is Hadamard and uniquely determined, 
if this is thought to be important.

Returning now to the simple harmonic oscillator, and to the unbounded case where
$M=(-\infty,\infty)$, we recognize that (22) corresponds exactly to
the minimum-energy ``vacuum''.  That in itself would not be surprising if our
S-J prescriptions had made some appeal to energy or to positive frequency, but
no amount of re-reading of the three characterizations given in the previous
section will reveal any such appeal, either overt or covert.  Why then did the
S-J vacuum of the oscillator turn out to coincide with its minimum energy state?

Evidently, our ``ground-state condition'' must have been responsible, but how?
In (0+1)-dimensions, $W$ will take the form
$$ 
      W(t,t') = f(t) \  \bar{f(t')}
$$ 
where, as we saw, $f$ will solve the oscillator equation of motion. 
Since $f(t)=\xi\sin\omega t+\eta\cos\omega t$ is the general solution, we will
have for the integeral of $f^2$,
$$ 
  \int\limits_{-\infty}^\infty f^2 
   = \int\limits_{-\infty}^\infty \xi^2 \sin^2 + \eta^2 \cos^2 + 2\xi\eta\sin\cos \ ,
$$ 
which will vanish iff $\xi^2+\eta^2=0$, given that the third term
oscillates and integrates to 0.  This in turn implies that 
$\xi=\pm i\eta$, whence (up to normalization)
$$ 
             f(t) = e^{\pm i\omega t}
$$ 
In this way, our ground-state condition singles out from all the solutions,
the two of  pure frequency.  Although that almost determines $W$, it still
leaves open the choice between between purely positive and purely negative frequency.
To grasp what resolves this final ambiguity, recall that $W$ was defined
to be the {\it\/positive\/} part of $i\Delta$.  
This implies that $\bra f|i\Delta|f\ket > 0$, 
which in turn rejects 
the negative frequency solution, 
leaving only $f(t)=e^{-i\omega t}$.
Perhaps then, it would be fair to summarize the explanation by saying that
``purity of $W$ $\implies$ purity of frequency'', 
while
``positivity of $W$ $\implies$ positivity of frequency''.

The explanation we have just gone through makes it clear that the S-J vacuum
will coincide with the minimum-energy   vacuum in any spacetime which
is stationary in the sense that it admits a Killing vector that is timelike
everywhere.  
In particular the usual and S-J vacua will coincide in $\Minkowski^4$.
(It is of course necessary that the minimum-energy vacuum exist itself!
For example, if $\phi$ is massless, one must exclude spacetimes like the static
cylinder, $\Reals\times S^3$, or for that matter $\Minkowski^2$.)

\bigskip
\QUESTION  What happens when the spacetime $M$ contains an ergoregion or horizon
(in other words when the Killing vector $\xi$ is spacelike in places)?

\bigskip
\REMARK  
From its definition, it is obvious that the S-J vacuum --- like any
vacuum --- is inherently global in nature.  Perhaps the simplest way to
make obvious that this could not have been avoided is to invoke a spacetime
which includes two static regions separated in time by a generic dynamical
region where a certain amount of ``particle creation'' will occur.  No
$W$ could then look like the locally defined vacuum in both static regions.
One knows after all, that ``particle'' is a nonlocal concept --- at least
for the kind of particle whose absence the word ``vacuum'' normally signifies!

\bigskip
\REMARK When $M$ is infinite, our prescription for $W$ need not converge.
Consider for example, the harmonic oscillator with $\omega=0$, more
commonly known as the free particle.  To what could $W$ possibly
converge?  If the limit existed and were unique then $W$ would have to
be time-translation invariant, so $W(t,t')=f(t-t')$, which seems
impossible, given our other conditions.  It might be interesting to see
exactly what goes wrong in this case when one takes $T$ to infinity in
$M=[-T,T]$.

\section{6.~For which $W$ does the entropy vanish?}                 
%
My final lecture at the Workshop was intended to cover entanglement entropy, but
time remained only for some brief remarks that mostly summarized material
already available in [16] [17].  
In addition however, they included a necessary
and sufficient algebraic condition for the entropy associated with $W$ to
vanish.  I will devote the present, final section to this ``purity criterion''
for $W$.

To any Wightman-tensor $W^{xy}$ there corresponds an entropy $S(W)$
defined by the sum, 
$$ 
    S = \sum\limits_\lambda  \lambda \log{|\lambda|}    \eqno(25) 
$$ 
where $\lambda$ runs over 
the solutions of the generalized eigenvalue equation,  
$$ 
    W f = i \, \lambda \; \Delta \, f    \eqno(26)
$$  
In the sum (25), each eigenvalue $\lambda$ must of course be
given its correct multiplicity, which can be done by treating as equal
any two covectors $f$ that differ by an element of $\ker{R}$, 
where $R$ is the real (equivalently the symmetric) part of 
$W=R+i\Delta/2$.

{\bf [} 
Thus, one treats $W$, $R$, and $\Delta$ as linear maps from
$V^*/\ker{R}$ to $V$, 
this being consistent because 
$W\ge0 \implies \ker{R}\subseteq\ker\Delta$, 
whence also
$\ker{R}\subseteq\ker{W}$.  
The multiplicity of a given eigenvalue $\lambda$ is then defined to be the dimension
of the subspace of $V^*/\ker{R}$ comprising the solutions $f$ of (26).
Here, $V^*$ is the dual space to $V$, and I have not bothered to
distinguish between $V$ and its complexification, $V\tensor\Complexes$,
in the formulas just written.
Notice also that 
$\ker{R}\subseteq\ker\Delta \implies \im\Delta\subseteq\im{R}$,
so that we can further regard $W$ as mapping $V^*/\ker{R}$ to $\im{R}=R[V^*]$.
In terms of a matrix representing $W$, this amounts to the following
more ``practical'' prescription:
diagonalize $W$ over the reals into $2\times2$ and $1\times1$ blocks and discard the zero blocks.
Notice finally, that although the positivity of $W$ entails
$\ker{R}\subseteq\ker\Delta$, 
the opposite inclusion can in priniciple fail. 
For such an $f$, namely one such that $\Delta f=0$ but $R f\not=0$, one
has $Wf=(R+i/2\Delta)f\not=0$, and
only $\lambda=\infty$ could satisfy (26).  This seems to imply
an infinite contribution to (25), 
and in fact an $f$ of this type, if present,  
would correspond to a ``purely classical component'' of $W$, 
which would indeed contribute an infinite entropy, 
the entropy of a classical gaussian probability
distribution being infinite unless one introduces some sort of cutoff.  
In a matrix representation, such an $f$ corresponds to a $1\times1$
block which is non-vanishing.
Before leaving the question of multiplicity, perhaps it's worthwhile to
mention yet another equivalent definition.  One can construe $W R^{-1}$
as an operator $T$ on $\im{R}$, and having done so, one sees by suitably
rearranging equation (26), that the eigenvalues of $T$ take the
form $\lambda/(\lambda-1/2)$.
{\bf ]}

Now let us regard $W$ as {\it\/pure\/} when $S(W)=0$.  It is clear from
(25) that this will be the case if and only if the eigenvalues
$\lambda$ are all either 0 or 1, which formally is expressed by the equality,
$(\Delta^{-1}W)^2=i(\Delta^{-1}W)$, or 
$$
   W \; \Delta^{-1} \; W = iW   \eqno(27)
$$
When $R$ and $\Delta$ have equal kernels, this equation is actually well
defined, even when (as will usually be the case) $\Delta$ is uninvertible. 
It could therefore be taken as our criterion of purity.  
More generally, though, one can take instead of (27) the rearrangement
$$
   W\;  R^{-1} \; W = 2 \, W    \eqno(28)
$$
which is well defined whenever $W\ge0$ (which it always is).
Both (27) and (28) possess equivalent forms involving
solely $R$ and $\Delta$, which respectively are
$$
    R \; \Delta^{-1} \; R = - \Delta/4  \quad  \hbox{ and } \quad \Delta \;  R^{-1} \; \Delta = \,-4 R
    \eqno(29\ a, b)
$$
It may be that (27)-(29) have not already appeared in the
literature in exactly this form.\footnote{$^\star$}
{Section 2.2 of [15] defines purity by saturation of a certain
 inequality of Cauchy-Schwarz type.  Very likely, it is provably equivalent to
 (28).}

\bigskip
\PROBLEM  Verify the claims about when (27)-(29) are well defined, and
verify that when defined, they are equivalent.  Prove more generally that (28)
[or (29b)] implies the other three.

\bigskip\PROBLEM  Show that the ``ground-state condition'', $W\Wbar=0$, implies 
 the purity condition,   \ $\Delta R^{-1} \Delta = -4 R \;$.  The S-J vacuum
 is therefore always pure.\footnote{$^\dagger$} 
{Mehdi Saravani has pointed out that the converse seems to hold as well: Every
 pure $W$ has the S-J form with respect to some metric on $V$.  In this sense,
 our ground-state condition is actually equivalent to purity.}


\bigskip
\noindent
In concluding, I would like to express my gratitude to
Niayesh Afshordi, 
Siavash Aslanbeigi, 
Michel Buck,
Fay Dowker, 
David Rideout, 
Mehdi Saravani, 
and 
Yasaman Yazdi
for contributing to the viewpoint
put forward herein.
Special thanks go to A.P.~Balachandran and other members of the audience
for asking how the Fok representation of Section 3 could be,
paradoxically, both irreducible and reducible at the same time.
And special thanks also to Marco Laudato for preparing the diagrams.

This research was supported in part by NSERC through grant RGPIN-418709-2012.
This research was supported in part by Perimeter Institute for Theoretical Physics.  
Research at Perimeter Institute is supported by the Government of Canada
through Industry Canada and by the Province of Ontario through the
Ministry of Research and Innovation.

\ReferencesBegin                             


\ref [1] Rafael D.~Sorkin, ``Quantum Mechanics as Quantum Measure Theory'', 
   \journaldata{Mod. Phys. Lett.~A}{9 {\rm (No.~33)}}{3119-3127}{1994}
   \eprint{gr-qc/9401003}
    \linebreak
   \eprint{http://www.pitp.ca/personal/rsorkin/some.papers/80.qmqmt.pdf}.
\sepref
{\'A}lvaro Mozota Frauca and Rafael Dolnick Sorkin,
``How to Measure the Quantum Measure''
\eprint{http://arxiv.org/abs/1610.02087}
\journaldata{Int. J. Theor. Phys.}{56(1)}{232-258}{2017}

\ref [2] J.B.~Hartle, ``Spacetime Quantum Mechanics and the Quantum Mechanics of Spacetime'',
 in B.~Julia and J.~Zinn-Justin (eds.),
 {\it Gravitation et Quantifications: Les Houches Summer School, session LVII, 1992}
 (Elsevier Science B.V. 1995), \linebreak
 \arxiv{gr-qc/9304006}

\ref [3] Luca Bombelli, Joohan Lee, David Meyer and Rafael D.~Sorkin, ``Spacetime as a Causal Set'', 
  \journaldata {Phys. Rev. Lett.}{59}{521-524}{1987}

\ref [4] Steven Johnston, {\it Quantum Fields on Causal Sets} PhD Thesis (Imperial College London, September 2010)
      arXiv:1010.5514 

\ref [5] R.D.~Sorkin and E.~Woolgar,             
  ``A Causal Order for Spacetimes with $C^0$ Lorentzian Metrics: 
    Proof of Compactness of the Space of Causal Curves'',
   \journaldata{Class. Quant. Grav.}{13}{1971-1994}{1996}
   \eprint{gr-qc/9508018}

\ref [6] Rafael D.~Sorkin and Yasaman K.~Yazdi, ``Entanglement Entropy in Causal Set Theory''
     arxiv:1611.10281

\ref [7] Rafael D.~Sorkin, {\it\/Notes on Stokes' Theorem\/}, \linebreak
     http://www.pitp.ca/personal/rsorkin/lecture.notes/stokes.theorem.pdf

\ref [8]  Rafael D.~Sorkin, ``Scalar Field Theory on a Causal Set in Histories form'' 
 \journaldata{Journal of Physics: Conf. Ser.}{306}{012017}{2011} 
 \arxiv{1107.0698},
 \eprint{http://www.pitp.ca/personal/rsorkin/some.papers/142.causet.dcf.pdf}

\ref [9] Urbashi Satpathi, Supurna Sinha and Rafael~D.~Sorkin,       
``A quantum diffusion law'' 
 \eprint{http://www.pitp.ca/personal/rsorkin/some.papers/158.brownian.pdf} \lbr
 \arxiv{1702.06273}
 (submitted)

\ref [10] Felix Finster, ``Definition of the Dirac Sea in the Presence of External Fields'', 
     \journaldata{Adv. Theor. Math. Phys.}{2}{963-985}{1998},
     see the discussion surrounding (3.9) and (3.10) 


\ref [11] {Niayesh Afshordi}, {Siavash Aslanbeigi} and {Rafael D.~Sorkin}
 ``A Distinguished Vacuum State for a Quantum Field in a Curved Spacetime: Formalism, Features, and Cosmology''
 \arxiv{1205.1296}
 \journaldata {JHEP}{2012 (8)}{137}{2012}

\ref [12] {Niayesh Afshordi}, {Michel Buck}, {Fay Dowker}, {David Rideout}, {Rafael D. Sorkin}, and {Yasaman K. Yazdi},
``A Ground State for the Causal Diamond in 2 Dimensions'' \linebreak
 \journaldata {JHEP}{2012 (10)}{088}{2012} 
 \arxiv{1207.7101}

\ref [13] Christopher J. Fewster and Rainer Verch, ``On a Recent
     Construction of `Vacuum-like' Quantum Field States in Curved Spacetime'',
  \journaldata{Classical and Quantum Gravity}{29(20)}{205017}{2012}
   arXiv:1206.1562 [math-ph]

\ref [14]  Christopher J Fewster, Rainer Verch, ``The Necessity of the Hadamard Condition'',
   Class. Quantum Grav. 30 (2013) 235027
   arXiv:1307.5242 [gr-qc]

\ref [15] Marcos Brum and Klaus Fredenhagen, ```Vacuum-like' Hadamard states for quantum fields on curved spacetimes'',
  arXiv:1307.0482 [gr-qc]

\ref [16] Rafael D. Sorkin, ``On the Entropy of the Vacuum Outside a Horizon'',  
  in B. Bertotti, F. de Felice and A. Pascolini (eds.),
  {\it Tenth International Conference on General Relativity and Gravitation (held Padova, 4-9 July, 1983), Contributed Papers}, 
  vol. II, pp. 734-736
  (Roma, Consiglio Nazionale Delle Ricerche, 1983),
  \linebreak
  \eprint{http://www.pitp.ca/personal/rsorkin/some.papers/31.padova.entropy.pdf}
 \linebreak
  \eprint{http://arxiv.org/abs/1402.3589};
\sepref
 Luca Bombelli, Rabinder K.~Koul, Joohan Lee and Rafael D.~Sorkin, 
``A Quantum Source of Entropy for Black Holes'', 
  \journaldata{Phys. Rev.~D}{34}{373-383}{1986}.

\ref [17] Rafael D.~Sorkin, ``Expressing entropy globally in terms of (4D) field-correlations'',
  \journaldata{J. Phys. Conf. Ser.}{ 484}{ 012004} {2014}
   (Proceedings of the Seventh International Conference on Gravitation and Cosmology [ICGC], held December 2011 in Goa, India)
  \linebreak
  \arxiv{1205.2953}  
  \linebreak
  \eprint{http://www.pitp.ca/personal/rsorkin/some.papers/143.s.from.w.pdf}

\ref [18] Mehdi Saravani, Rafael D.~Sorkin, and Yasaman K.~Yazdi ``Spacetime Entanglement Entropy in 1+1 Dimensions''
  \journaldata{Class. Quantum Grav.}{31}{214006}{2014} 
   available at http://stacks.iop.org/0264-9381/31/214006,
  \arxiv{1205.2953}
  \linebreak
  \eprint{http://www.pitp.ca/personal/rsorkin/some.papers/148.pdf}


\end                                         


(prog1 'now-outlining
  (Outline* 
     "\f"                   1
      "
      "
      "
      "
      "\\Abstract"          1
      "\\section"           1
      "\\subsection"        2
      "\\appendix"          1       ; still needed?
      "\\ReferencesBegin"   1
      "
      "\\ref "              2
      "\\end